\documentclass[aps,twocolumn,floats,prd,nofootinbib,showpacs,showkeys]{revtex4}

\usepackage[dvips]{graphicx} 
\usepackage{epsfig,amsmath}
\usepackage{amssymb}

\usepackage{rotate}
\usepackage{color}
\usepackage{bm}

\DeclareFontFamily{OT1}{pzc}{}
\DeclareFontShape{OT1}{pzc}{m}{it}%
            {<-> s * [1.10] pzcmi7t}{}
\DeclareMathAlphabet{\mathscr}{OT1}{pzc}%
                                {m}{it}

\def\nhat{\hat{n}}

\newcommand{\be}{\begin{equation}}
\newcommand{\ee}{\end{equation}}
\newcommand{\bea}{\begin{eqnarray}}
\newcommand{\eea}{\end{eqnarray}}
\def\ba#1\ea{\begin{align}#1\end{align}}

\newcommand{\refeq}[1]{Eq.~(\ref{eq:#1})}          
\newcommand{\refeqs}[2]{Eqs.~(\ref{eq:#1})--(\ref{eq:#2})}          
\newcommand{\reffigs}[2]{Figs.~\ref{fig:#1}--\ref{fig:#2}}          
\newcommand{\reffig}[1]{Fig.~\ref{fig:#1}}          
\newcommand{\refsec}[1]{Sec.~\ref{sec:#1}}
\newcommand{\refapp}[1]{App.~\ref{app:#1}}
\newcommand{\vs}{\nonumber\\}

\newcommand{\bga}{\begin{gathered}}
\newcommand{\ega}{\end{gathered}}
\newcommand{\beqa}{\begin{eqnarray}}
\newcommand{\eeqa}{\end{eqnarray}}

\newcommand{\degree}{\ensuremath{^\circ}}

\newcommand{\id}{{\rm d}}
\newcommand{\kB}{k_{\rm B}}

\definecolor{RedWine}{rgb}{0.743,0,0}
\definecolor{RoyalBlue}{rgb}{0.25,.41,.88}

\begin{document}

\title{The effect of aberration on partial-sky measurements of
     the cosmic microwave background temperature power spectrum}

\author{Donghui Jeong, Jens Chluba, Liang Dai, Marc
     Kamionkowski, and Xin Wang}
\affiliation{Department of Physics and Astronomy, Johns
     Hopkins University, 3400 N.\ Charles St., Baltimore, MD 21218}

\begin{abstract}
 Our motion relative to the cosmic-microwave-background (CMB)
 rest frame deflects light rays giving rise to shifts as large as
 $\ell \to \ell(1\pm\beta)$, where $\beta=0.00123$ is our velocity
 (in units of the speed of light)  on measurements 
 CMB fluctuations.  
 Here we present a novel harmonic-space approach to this
 CMB aberration that improves upon prior work by allowing us to
 (i) go to higher orders in $\beta$, thus extending the validity
 of the analysis to measurements at $\ell\gtrsim \beta^{-1}\simeq
 800$; and (ii) treat the effects of window functions and
 pixelization in a more accurate and computationally efficient
 manner.  We calculate precisely the magnitude of the systematic
 bias in the power spectrum inferred from the partial sky,
 and show that aberration shifts the multipole moment 
 by $\Delta\ell/\ell\simeq\beta\left<\cos\theta\right>$,
 with $\left<\cos\theta\right>$ averaged over the survey footprint.
 Such a shift, if ignored, would bias the measurement of the sound-horizon size
 $\theta_{\mathrm *}$ at the $0.01\%$-level, which is comparable to the 
 measurement uncertainties of Planck.
 The bias can then propagate into cosmological parameters such as 
 the angular-diameter distance, Hubble parameter and dark-energy 
 equation of state.
 We study the effect of aberration for
 current Planck, South Pole
 Telescope (SPT) and Atacama Cosmology Telescope (ACT) data and
 show that the bias cannot be neglected.  
 We suggest that the small tension between Planck, ACT, and SPT may
 be due partially to aberration.  An Appendix shows
 how the near constancy of the full-sky power spectrum under
 aberration follows from unitarity of the aberration kernel.
\end{abstract}

\date{\today}

\maketitle

\section{Introduction}
Cosmic microwave background (CMB) radiation is very nearly
isotropic in all directions with temperature of
$\bar{T}=2.7260\pm0.0013~\mathrm{K}$
\cite{Fixsen:1996nj}.  There is, however, 
a temperature dipole ($\ell=1$) in the CMB \cite{dipole} that is
now measured to have an amplitude
$\Delta T\simeq 3.355\pm 0.008~\mathrm{mK}$ along the direction 
$(l,b)=(263.99\degree\pm0.14\degree,48.26\degree\pm0.03\degree)$
in Galactic coordinates \cite{Hinshaw:2008kr}.
Attributing the full dipole anisotropy to our motion with respect 
to the CMB rest frame suggests a speed of
$v=369\pm0.9~\mathrm{km/s}$, or $\beta \equiv v/c = 0.00123$.

The CMB temperature also shows anisotropies, at the level of
tens of $\mu$K, of primordial origin that are statistically
isotropic in the rest frame of the CMB.
However, our motion with respect to the CMB rest frame
causes light aberration, a coherent modulation of the observed
angular position of CMB photons from the original direction in
the CMB rest frame.  This leads to a dipolar departure from
statistical isotropy:  hot/cold spots observed in the direction
of our motion appear to be smaller by a factor $\simeq(1- \beta)$
than in the CMB rest frame, and {\it vice versa} in the opposite
direction.  The Planck collaboration recently detected this
aberration \cite{Aghanim:2013suk} and confirmed that it is
consistent with the velocity derived from the dipole
anisotropy.

Most of the recent literature on light aberration traces back to
Ref.~\cite{Challinor:2002zh} which considered the effects of
aberration to $O(\beta^2)$ in full-sky CMB maps, concluding that
aberration would lead to a correction $O(\beta^2)\simeq10^{-6}$ to
the values of cosmological parameters inferred by a full-sky CMB
map.  Refs.~\cite{Kosowsky:2010jm,Amendola:2010ty} considered
the detectability of the off-diagonal correlations of temperature
anisotropies induced by aberration, and Ref.~\cite{Burles:2006xf} 
considered the effect of aberration on the angular scale of the 
acoustic peaks in the CMB power spectrum. 
Ref.~\cite{Pereira:2010dn} showed that the
cancellations between the forward and backward directions in a
full-sky map that lead to an $O(\beta^2)$ correction to the power
spectrum do not arise in a partial-sky map.  In this case, the
effects of aberration on the inferred power spectrum appear at
$O(\beta)$, and are thus considerably larger.
Ref.~\cite{Pereira:2010dn} thus noted the significance of this
effect for an experiment like Planck, which maps the full sky
but masks out in the analysis portions of the sky obscured by
foregrounds.  The determination of cosmological parameters from
these temperature anisotropies \cite{Jungman:1995av} may
therefore be affected at $O(\beta)$.

It has also been noted that the analytic approach of
Ref.~\cite{Challinor:2002zh}, which works to $O(\beta^2)$,
breaks down for multipole moments $\ell\gtrsim \beta^{-1}\simeq
800$ \citep[e.g., see][]{Challinor:2002zh, Chluba:2011zh}.  Thus, for experiments like Planck \cite{Ade:2013zuv}, the South Pole
Telescope (SPT) \cite{Story:2012wx,Hou:2012xq}, and the Atacama
Cosmology Telescope (ACT) \cite{Das:2013zf,Sievers:2013ica},
which make high-angular-resolution precision
measurements of CMB fluctuations, something more must be done.
Refs.~\cite{Notari:2011sb,Yoho:2012am} thus proposed to deal
with this issue by ``de-boosting'' the CMB map in real-space;
i.e., using the magnitude and direction of the temperature dipole
to Lorentz transform the observed CMB temperature to the rest
frame.  As we show below, however, this real-space de-boosting
does not easily account for effects associated with pixelization
and a finite window function, issues that arise from the
Lorentz transformation of the solid angle from the observer
frame to the rest frame.

Here we employ a harmonic-space approach to CMB aberration.
We adopt the recursive calculation of Ref.~\cite{Chluba:2011zh}
to include the effects of aberration to higher orders in $\beta$
and thus to treat maps with very high resolution.  It also
provides, as we show below (and argued in
Ref.~\cite{Chluba:2011zh}), a far more effective and
computationally efficient way to include the effects of
pixelization and window function.  To be a bit more precise, we
combine the real- and harmonic-space approaches. We first
generate Gaussian random realizations for a given power spectrum
in harmonic space and then include the effects of aberration by
transforming the $a_{\ell m}$s with the harmonic-space
aberration kernel from Ref.~\cite{Chluba:2011zh}.  We then
transform these harmonic coefficients to real space to apply the
masks by multiplying the masking function ($1$ for the observed
pixel, $0$ otherwise). Finally, we measure the angular power
spectrum from the masked real-space map.  We repeat this
procedure many times to study systematic changes in the
resulting angular power spectrum and compare the resulting
angular power spectrum to the masked angular power spectrum
without aberration. As we show below, this approach accounts for
mask window-function effects more accurately with less computational
effort.

We then apply these calculations to determine the effects of
aberration on SPT and ACT (and also Planck).  As we will see,
the magnitude of the aberration correction to the ACT/SPT power
spectra are closer to $\simeq 1\%$, rather than $O(\beta)
\simeq 0.1\%$.  This is because the power spectrum scales roughly as
$C_\ell \propto \ell^{-7}$ for $1000 \lesssim \ell \lesssim
3000$ and because aberration leads (in the forward/backward
direction) to a re-scaling $\ell \to \ell(1\pm\beta)$.  The
fractional change to the power spectrum is thus $(\Delta
C_\ell/C_\ell) \simeq 7 \beta\simeq 0.01$, and even larger in
regimes where the acoustic oscillations increase $\partial
C_\ell/\partial \ell$.  This correction is thus comparable in
magnitude to the statistical error in these experiments and is
thus a systematic correction that {\it must} be taken into
account in measurement.  It is thus imperative to perform the
correction carefully, as we do below.

This paper is organized as follows. We begin in \refsec{fullsky}
by reviewing the effect of aberration on the full sky.  Then, in
\refsec{simulation}, we discuss simulations of aberration on the
full sky including the effects of pixelization and the window
function.  \refsec{cutsky} then discusses measurements of the
power spectrum on the cut sky.  After considering some
illustrative examples, we calculate explicitly the effects of
aberration on the power spectra inferred in Planck, ACT (both
the equatorial and southern surveys), and SPT. We conclude in
\refsec{conclusion}.  An Appendix discusses the unitarity of the
aberration kernel and shows why the effects of aberration on the
full-sky power spectrum remain small even at $\ell \gtrsim 800$.

\section{Aberration on the full sky}\label{sec:fullsky}
\subsection{Basics: CMB aberration}
Let us denote the 4-momentum of a CMB photon by $p^\mu = (E,
\mathbf{p})$ in the CMB rest frame and ${p^\mu}' =
(E',\mathbf{p}')$ in the observer's frame. For simplicity, let
us align the $z$ axis with the direction of the observer's
motion with respect to the CMB rest frame. Then, the two
4-momenta are related by a Lorentz boost,
\be
\left(\begin{array}{c}
E \\ p_x \\ p_y \\ p_z\\
\end{array}\right)
=
\left(\begin{array}{cccc}
\gamma & 0 & 0 & \gamma\beta\\
0 & 1 & 0 & 0\\
0 & 0 & 1 & 0 \\
\gamma\beta & 0  &0 & \gamma\\
\end{array}\right)
\left(\begin{array}{c}
E' \\ p_x' \\ p_y' \\ p_z'
\end{array}\right).
\ee
Here, $\beta = v/c$, and $\gamma = 1/\sqrt{1-\beta^2}$.
Consider a CMB photon seen in the CMB rest frame in the direction 
$-\mathbf{\hat{p}} = \mathbf\nhat = (\sin\theta,0,\cos\theta)$. 
Then, we observe the photon from the direction 
$-\mathbf{\hat{p}}'=\mathbf\nhat' = (\sin\theta',0,\cos\theta')$ where 
\be
\cos\theta' = \frac{\cos\theta + \beta}{1+\beta\cos\theta}.
\label{eq:angleT}
\ee
Under this transformation, the solid-angle element transforms as 
\be
\id\Omega' = 
\frac{\id\Omega}{\gamma^2(1+\beta\cos\theta)^2},
\label{eq:jacobian}
\ee
which means that the observed solid angle covered by a bundle of
CMB photons is different than that in the CMB rest frame.
As a result,
the angular separation $\Delta\theta$ in the CMB rest frame is observed 
to be 
\be
\Delta\theta' 
= \frac{\Delta\theta}{\gamma (1+\beta\cos\theta)}
\simeq \Delta\theta\left[1-\beta\cos\theta + \mathcal{O}(\beta^2)\right].
\label{eq:Deltatheta_aberration}
\ee
That is,
when considering typical cold or hot spots 
of $\simeq1\deg$, those spots shrink toward the direction of our motion 
($\theta=0$) and expand toward the opposite direction ($\theta=\pi$).
Therefore, the temperature (and polarization) anisotropies show a dipolar
distortion in their shape, which we call \textit{aberration}.

The specific intensities in the CMB rest frame and
observer frame are related by
\be
I_{\nu'}(\nhat') = \frac{\nu'^3}{\nu^3} I_{\nu}(\nhat),
\ee
with $\nu=\gamma(1-\beta\cos\theta')\nu'$ and $\nhat'$ and $\nhat$ 
being related by \refeq{angleT}.
If the specific intensity in the CMB rest frame is a Planck
function with temperature $T(\nhat)=\bar{T}[1+\Delta_T(\nhat)]$,
\be
I_{\nu}(\nhat) = \frac{2h\nu^3}{c^2}
\left[\exp\left(\frac{h\nu}{\kB T(\nhat)}\right)-1\right]^{-1},
\ee
then the intensity in the observed frame is given by
\be
     I_{\nu'}(\nhat') = \frac{2h\nu'^3}{c^2}
     \left[\exp\left(\frac{h\nu'\gamma(1-\beta\mu')}{\kB
     T(\nhat)}\right)-1\right]^{-1},
\ee
which is, again, a Planck function with observed temperature
\be
     T_{\rm obs}(\nhat') = \frac{T(\nhat)}{\gamma(1-\beta
     \mu')}.
\label{eq:DT}
\ee
This is the key equation of aberration which relates the
observed temperature at direction $\nhat'$ to the
intrinsic CMB temperature $T(\nhat)$.

\subsection{Aberration in harmonic space}
\label{sec:kernel}

In spherical-harmonic space, the aberration in \refeq{DT} is
written as a linear transformation\footnote{There is a
correction that arises from the fact that the mean temperature
of the boosted map is different than that of the original map
\cite{Chluba:2011zh}.  However, except for the temperature dipole
which is linear in $\beta$, 
this correction is tiny, $O(\beta^2)$, and so we neglect it below.}
\cite{Challinor:2002zh},
\be
     a^{(\rm obs)}_{\ell m}(\boldsymbol\beta) =  \sum_{\ell'm'}
     \mathcal{K}_{\ell m}^{\ell'm'}(\boldsymbol\beta)
     a_{\ell'm'},
\label{eq:def_kernel}
\ee
of the spherical-harmonic coefficients.  Here $\mathcal{K}_{\ell
m}^{\ell'm'}$ is the aberration kernel which depends on the
amplitude $\beta$ and direction $\hat{\beta}$ of the
observer's velocity in the CMB rest frame.  We indicate the
observed spherical-harmonic coefficients in the moving
observer's frame by a subscript $(\rm obs)$.  The kernel,
obtained from \refeq{DT}, is
\ba
\mathcal{K}^{\ell'm'}_{\ell m}({\boldsymbol\beta}) = 
\int \frac{\id^2\Omega'}{\gamma(1-\boldsymbol\beta\cdot\mathbf{\hat n'})}
Y_{\ell'm'}(\nhat) Y_{\ell m}^*(\nhat').
\label{eq:kernel-expr}
\ea
When we choose the coordinate system so that the observer's moving
direction is aligned to the pole ($\theta=0$), the azimuthal symmetry
allows the kernel to be simplified to
\be
\mathcal{K}_{\ell m}^{\ell'm'}(\beta \mathbf{\hat{z}}) 
\equiv \mathcal{K}_{\ell m}^{\ell'm}(\beta) \delta_{mm'},
\label{eq:kappasym}
\ee
where
\ba
\mathcal{K}^{\ell'm}_{\ell m}(\beta)
=&
2\pi
\mathcal{N}_{\ell m}\mathcal{N}_{\ell'm}
\int_{-1}^1 \frac{P_{\ell'}^m(\mu)P_{\ell}^m(\mu')}{\gamma(1-\beta\mu')}
{\rm d}\mu',
\label{eq:kappa}
\ea
in terms of associate Legendre polynomials $P_\ell^m(\mu)$ and 
the normalization factor
\be
\mathcal{N}_{\ell m}
=
\sqrt{\frac{(2\ell+1)}{4\pi}
\frac{(\ell-m)!}{(\ell+m)!}},
\ee
of the spherical harmonics.  
We evaluate the aberration kernel $\mathcal{K}_{\ell' m}^{\ell m}$
to higher orders in $\beta$
by using a recursion relation 
described in \cite{Chluba:2011zh}. Note that
the kernel satisfies the relations,
\ba
\mathcal{K}_{\ell m}^{\ell'm}(\beta)
=&
(-1)^{\ell+\ell'}
\mathcal{K}_{\ell'm}^{\ell m}(\beta),
\\
\mathcal{K}_{\ell-m}^{\ell'-m}(\beta)
=&
\mathcal{K}_{\ell m}^{\ell'm}(\beta),
\ea
so it suffices to calculate only a quarter of all the matrix elements.

The kernel $\mathcal{K}_{\ell'm}^{\ell m}$ quantifies how much
power is transferred from intrinsic CMB multipole coefficients
$a_{\ell'm}$ to $a_{\ell m}^{(\rm obs)}$.  \refeq{kappa}
therefore says that (1) aberration does not alter the
component of angular momentum in the direction $\hat{\beta}$ of
the moving observer; and (2) power transfer from $a_{\ell m}$ to
$a_{\ell+\Delta \ell, m}^{(\rm obs)}$ is most efficient for 
$\Delta \ell \lesssim \beta\ell$ and sharply decays for larger $\Delta \ell$.
For details of the aberration kernel, including 
the $\Delta\ell$ and $m$ dependence, we refer the reader to 
Ref.~\cite{Chluba:2011zh}.

\subsection{Aberrated CMB two-point functions}
Statistical isotropy dictates that the spherical-harmonic
coefficients $a_{\ell m}$ of the CMB map, in its rest frame, are
statistically independent.  However, aberration induces
correlations between different observed spherical-harmonic
coefficients $a_{\ell m}^{({\rm obs})}$.

Here, and throughout, we define $m$ with respect to the moving
direction, so that \refeq{kappasym} holds.  Then, the two-point
correlator of temperature anisotropies becomes
\ba
&
\left<
a_{\ell_1m_1}^{({\rm obs})}
a_{\ell_2m_2}^{({\rm obs})\,*}
\right>
\vs
=&
\sum_{\ell_1'\ell_2'm_1'm_2'}
\mathcal{K}_{\ell_1m_1}^{\ell_1'm_1'}(\beta)
\mathcal{K}_{\ell_2m_2}^{\ell_2'm_2'}(\beta)
\left<
a_{\ell_1'm_1'}
a_{\ell_2'm_2'}^*
\right>
\vs
=&
\sum_{\ell_1'\ell_2'm_1'm_2'}
\mathcal{K}_{\ell_1m_1}^{\ell_1'm_1}(\beta)\delta_{m_1m_1'}
\mathcal{K}_{\ell_2m_2}^{\ell_2'm_2}(\beta)\delta_{m_2m_2'}
C_{\ell_1'} 
\delta_{\ell_1'\ell_2'}
\delta_{m_1'm_2'}
\vs
=&
\sum_{\ell'}
\mathcal{K}_{\ell_1m_1}^{\ell'm_1}(\beta)
\mathcal{K}_{\ell_2m_2}^{\ell'm_2}(\beta)
C_{\ell'} 
\delta_{m_1m_2}.
\label{eq:two-pt-correlator}
\ea

\section{Simulating the effects of aberration}
\label{sec:simulation}
\begin{figure}[htbp]
\includegraphics[width=0.52\textwidth]{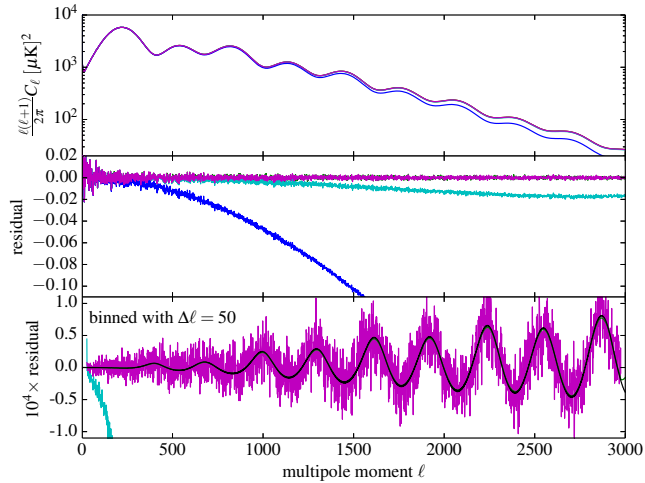}
\caption{Angular power spectra and their residual $\Delta C_\ell/C_\ell$
     for simulated full-sky temperature maps. The angular
     power spectra are averaged over 500 Gaussian realizations
     after taking the effect of our motion into account
     using real-space boosting (\textit{blue}) or
     harmonic-space boosting (\textit{magenta}).
     The central panel shows the residual
     relative to the input theoretical power spectrum,
     and the bottom panel shows the residual with respect to the
     power spectrum estimated from the un-aberrated Gaussian
     temperature map.  Thus, the cosmic variance cancels out in
     the bottom panel.
     Further, to reduce the scatter to the $10^{-5}$ level, 
     we take a moving average with width $\Delta\ell=50$ in the bottom panel.
     The black solid curve in the bottom panel shows the residual 
	  predicted from the aberration kernel and input
          theoretical power spectrum,
     and the green solid curve, which lies behind the black curve, 
     in the bottom panel shows the analytical 
     approximation in \refeq{DClovCl_second} with 
     $\left<\cos^2\theta\right>=1/3$.
     While harmonic-space boosting 
	  yields the aberrated angular power spectrum consistent with 
     the theory prediction, real-space
     boosting shows strong power suppression for
     $\ell\gtrsim300$.  Deconvolution of the window function
     (\textit{cyan}) restores somewhat the power on larger
     angular scales ($\ell \lesssim 500$), to within a percent level,
     but the deviation remains as large as $\simeq 2\%$ at $\ell\simeq 2500$
     because of aliasing and imperfect knowledge of the exact pixel
     window function in {\sc Healpix}.}
\label{fig:simCl_fullsky}
\end{figure}

In order to simulate the combined effect of aberration and sky
masking (the subject of the following Section) we must first
generate the temperature map in the observed frame.  A Gaussian
realization of the temperature map in the CMB rest frame is
easily obtained from a given angular power spectrum.  Once a map
of the temperature in the CMB rest frame is given, the effects
of aberration can be added in both real and harmonic space.

In real space, we add aberration by applying \refeq{DT} to the
simulated Gaussian temperature map. As the simulated map is
given in discretized pixels, we aberrate it in a pixel-by-pixel
manner.  We find it more convenient to apply \refeq{DT} backward
rather than forward.  Namely, for a pixel at location $\nhat'$
in the observed frame, we find $\Delta_T(\nhat)$, with $\nhat$
given by \refeq{angleT}, from the temperature in the CMB rest
frame.  Then, applying the $\mu'$-dependent modulation provides
the effect of aberration.  If we had alternatively started with
pixels in the CMB rest frame, we would have additionally needed
to take into account the changes [\refeq{jacobian}] in the solid
angle of each pixel.  This simple procedure was used in
Ref.~\cite{Chluba:2011zh} to illustrate the  effect of our
motion on the CMB sky.

The harmonic-space approach makes use of the aberration kernel
$\mathcal{K}_{\ell m}^{\ell'm}(\beta)$ calculated in
\refsec{kernel}. 
To ensure the convergence of the harmonic-space kernel,
we choose rather stringent parameters in the kernel calculation: 
mode mixing to $|\ell'-\ell|=10$ and terms to $\beta^{40}$,
or $\Delta \ell=10$ and $k_{\rm max}=20$ in terms of parameters 
in Ref. \cite{Chluba:2011zh}. With these settings the effect of 
aberration can be modeled quasi-exactly up to $\ell\simeq 4000$.
Once the kernel is calculated for a given observer velocity $\beta$,
we calculate the $a_{\ell'm}^{(\rm obs)}$ 
from the simulated Gaussian map $a_{\ell m}$ in harmonic space.

We generated 500 Gaussian temperature maps using the WMAP7
best-fit power spectrum \cite{Komatsu:2010fb} using {\sc
Healpix} \cite{Gorski:2004by}\footnote{{\sf http://healpix.jpl.nasa.gov}} with $N_{\rm side}=2048$ ($L_{\rm max}=4000$).
We then simulate aberration by using both real-space 
and harmonic-space boosting. The resulting power spectra averaged
over the realizations and their residuals are shown in
\reffig{simCl_fullsky}.
The upper panel shows the averaged angular power spectrum. 
On the full sky, an accurate simulation of the effect of aberration
must yield a temperature power spectrum close to the non-aberrated 
one \cite{Challinor:2002zh}.
While the angular power spectrum from the aberrated map using the
harmonic-space kernel (magenta curve) is consistent with the
input power spectrum, the angular power spectrum obtained with
real-space boosting (blue curve) shows a significant power
suppression on smaller angular scales.  We facilitate the
comparison better by dividing the aberrated angular power
spectrum by the input power spectrum in the central panel. 
For $N_{\rm side}=2048$, real-space boosting causes a
suppression of the power spectrum by $5\%$ at $\ell\simeq 1000$
and the suppression becomes larger on smaller scales (larger $\ell$). 

This suppression of power in real-space boosting was also found
in Refs.~\cite{Notari:2011sb,Yoho:2012am}.  There, to correct
for aberration, the authors inverted \refeq{DT} in real space to
recover the intrinsic temperature map from the observed
$\Delta_T'(\nhat')$. In their numerical study of real-space
de-boosting, Ref.~\cite{Yoho:2012am} report the suppression in
the recovered angular power spectrum relative to the input power
spectrum, and find that a {\sc Healpix} resolution of $N_{\rm
side}=8192$ or greater is required to achieve $1\%$ accuracy in
the recovered angular power spectrum to $\ell=2000$.  Performing
this analysis with such a high resolution is computationally
intensive and impractical with current computational power.
Harmonic-space boosting thus provides significant 
computational advantages.

\subsubsection{Effect of the pixel window function}

The power suppression in real-space boosting occurs when
applying the coordinate transformation from the moving frame to
the CMB rest frame: the pixel center in the observed frame
is not necessarily mapped into the pixel center in the CMB 
rest frame, and {\it vice versa}.  Therefore, aberration of
a given temperature map in the CMB rest frame always involves
interpolation to calculate the temperature anisotropy at the
mapped off-center point.

A generic interpolation scheme is formulated as the convolution
between pixelized temperature anisotropies and the interpolating
window function $W$:
\be
\Delta_T(\nhat) = \sum_{i} W(\nhat-\nhat_i)\Delta_T^{\rm(pixel)}(\nhat_i).
\label{eq:interpol}
\ee
Here, $\nhat_i$ is the angular position of $i$-th pixel-center.
The interpolating window function $W$ depends on the
interpolation scheme, but, in general, the shape and size of the
discretized  pixel are most important. For example, for the
nearest-grid-point (NGP) interpolation, where $\Delta_T(\nhat)$
takes the value at the nearest pixel point, $W$ is the pixel
shape itself [$1/$(Area of pixel) for the  
points inside of pixel, and $0$ otherwise], and for the
cloud-in-cell (CIC) interpolation, where $\Delta_T(\nhat)$ is
given by the weighted---weighting factor is proportional to the
proximity---sum of the four nearby $\Delta_T(\nhat_i)$'s, the
window function $W$ is given by the convolution of two pixel
shapes.  In harmonic space, convolution in \refeq{interpol}
implies that the resulting harmonic coefficients are given by
the multiplication of temperature anisotropies and the
interpolating window function $w_{\ell m}$:
\be
a_{\ell m}^{\rm (pixel)} = w_{\ell m}^{\rm (pixel)}a_{\ell m}.
\ee
Since it is localized in real space, the interpolating window
function is unity for $\ell \lesssim \pi/(\Delta\theta)$ but
decreases for small scales $\ell \gtrsim \pi/(\Delta\theta)$
where $\Delta\theta$ is the angular resolution.  Therefore, the
angular power spectrum of the aberrated map generated with
real-space boosting is expected to show power suppression on
smaller scales according to the shape of the window function $W$
of the interpolation.

Now that we know that power suppression is due to the
interpolation window function, we may remedy the situation by
de-convolving the window function in harmonic space.  As we
used the cloud-in-cell scheme, the
interpolating window function is given by 
\be
W_\ell\equiv \left[w_\ell{\rm (pixel)}\right]^2,
\ee
where $w_\ell \simeq \mathrm{sinc}(\ell\Delta\theta/2\pi)$ is
the pixel window function of {\sc Healpix}. Then, we deconvolve
the aberrated power spectrum as
\be
C_\ell^{\rm(deconvolve)} =W_\ell^{-2} \,C_\ell^{\rm(aberration)}.
\ee
The cyan curves in \reffig{simCl_fullsky} show the result of
deconvolution by dividing the angular power spectrum
from real-space boosting by the harmonic transform of the
{\sc Healpix} window function.  The de-convolved angular power
spectrum is somewhat improved relative to the severe power
suppression at multipoles $\ell\lesssim500$, but it
still deviates at $\lesssim2\%$ on smaller scales ($\ell\gtrsim2000$)
from the input power spectrum.  We attribute these residuals to
the imperfect knowledge of the {\sc Healpix} window  
function. In fact, because each pixel in {\sc Healpix} is not
identical, it is almost impossible to perfectly calculate the
pixel window function.  In addition, aliasing due to
the finite pixelization may also hamper the deconvolution.

To summarize, since the coordinate transformation \refeq{angleT}
is defined in real space, simulating the effect of aberration
may be easier in real space than in harmonic space, where
one has to separately calculate the kernels.  In practice,
however, we find the harmonic-space approach superior because
the finite resolution and associated pixel window function
plagues the real-space simulations.  Therefore, in the following
Section, we simulate aberration in harmonic space.

\subsubsection{The effect of aberration effect on the full-sky
angular power spectrum}
Although only up to $0.01\%$, the aberrated angular power spectrum do show 
systematic residual $\Delta C_\ell/C_\ell$ compared to the un-aberrated one.
The bottom panel of \reffig{simCl_fullsky} shows detailed shape of the 
power spectrum residual. To capture such a small residual from $500$ 
simulations, we first divide aberrated power spectrum by the 
power spectrum estimated from the un-aberrated Gaussian temperature map
so that the cosmic variance cancels out. Then, we further reduce the
scatter by taking a moving average with width $\Delta \ell = 50$
for harmonic-space boosting (magenta curve) and real-space
boosting (cyan curve).
For comparison, the black solid curve shows the theoretical
prediction calculated 
from the input power spectrum and aberration kernel with 
\refeq{two-pt-correlator}. Again, the power spectrum from harmonic-space 
boosting lies right on top of the theory prediction, while
real-space boosting, even after the deconvolution, fails to catch up 
with the correct result.

We observe that the residual due to aberration increases toward the smaller 
angular scales and oscillates roughly, but not exactly, 
in-phase with the second derivative of the angular power spectrum: 
$d^2C_\ell/d\ell^2$. This can be understood from unitarity as well as 
the symmetric shape of the aberration kernel.
We discuss  in \refapp{unitarity} and in the discussion
around \refeq{DClovCl_second} in the next Section
the analytical approximation of the residual which is shown as a
green line in the bottom panel of \reffig{simCl_fullsky}.
Note that the analytical approximation (green curve) is so accurate that it 
is hard to distinguish from the exact theory calculation (black curve).

\section{Aberration on a cut sky}\label{sec:cutsky}

In the previous Section, we studied the effects of aberration on
full-sky CMB temperature maps.  In reality, however, we never
perform analyses on the full sky.  Planck masks out regions with
bright foregrounds, and more importantly, suborbital experiments
like SPT and ACT are restricted to a small patch of sky.
In particular, high-resolution surveys like SPT and ACT 
measure the CMB power spectrum at $\ell\gg\beta^{-1}$ 
where aberration might be important.
In this Section, we demonstrate that aberration
introduces a systematic bias on the small-scale CMB 
power spectrum if the sky mask is significantly asymmetric
with respect to the equatorial plane of aberration.

Using the harmonic-space boosting approach, outlined in
Sec.~\ref{sec:simulation}, we simulate both aberrated and
unaberrated maps of the CMB assuming
$\beta=10^{-3}$ and measure the angular power spectrum only from 
the masked regions in real space to
examine the fractional change in
the cut-sky power spectrum induced by aberration%
\footnote{One can also convolve the two-point correlator in 
\refeq{two-pt-correlator} with the masking window function in harmonic space.
However, we find that this procedure takes an impractically long
time as it requires
two convolutions each of which involves multi-dimensional integration.}.
As we will see, the effect of aberration on the cut-sky power spectrum depends
sensitively on the \textit{shape} and \textit{location} of the masked region. 
When the masked region is symmetrically distributed in the forward/backward 
direction with respect to our peculiar velocity, the effect of aberration
is essentially the same as in the full-sky case. On the other hand,
when the masked region contains more in the forward (backward) direction,
the aberrated power spectrum is enhanced (suppressed) compared to the 
un-aberrated power spectrum. Therefore,
ACT and SPT, both of which are aimed largely in the backward direction, 
suffer from power suppression as large as $1\%$,
and the high-frequency channel of Planck is also affected by aberration at
the level comparable to the cosmic-variance error on smaller scales.
A bias at this level can result in
discrepancies between these CMB experiments if aberration is not
accounted for, as we demonstrate below.

\subsection{Illustrative examples}

\begin{figure*}[htbp]
\includegraphics[width=0.50\textwidth]{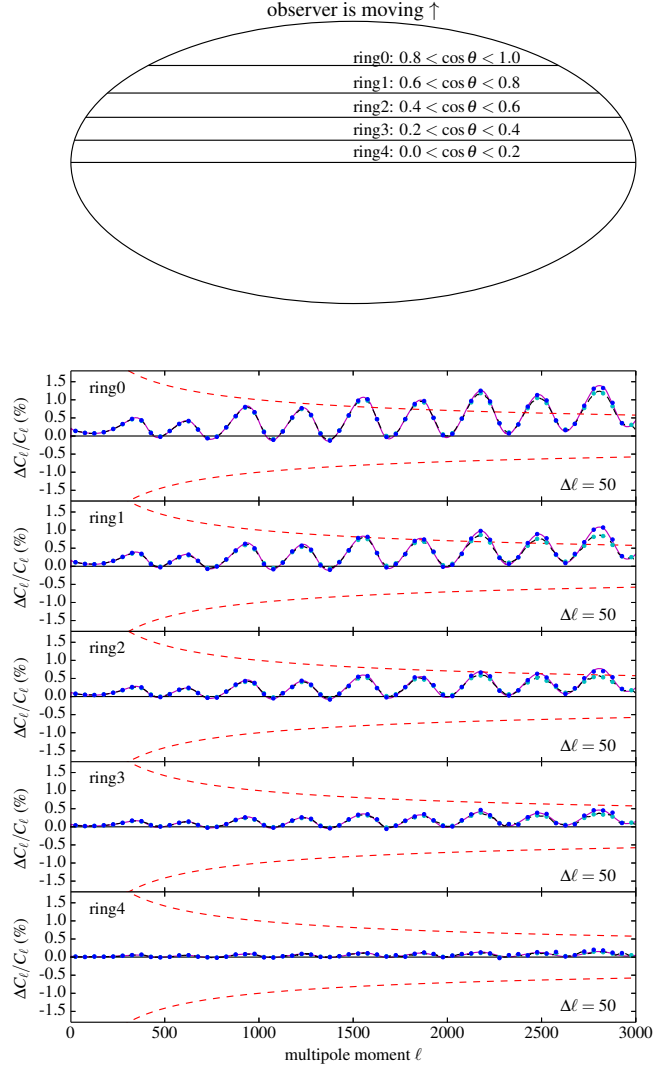}
\caption{\textit{Top:} shape of five toy `ring' surveys
     defined by  $0.8<\cos\theta<1$ ({\sf ring0}),
     $0.6<\cos\theta<0.8$ ({\sf ring1}), $0.4<\cos\theta<0.6$ ({\sf ring2}),
     $0.2<\cos\theta<0.4$ ({\sf ring3}), and $0<\cos\theta<0.2$ ({\sf ring4}), 
     so that each survey region has $f_{\rm sky}=0.1$.  
     The observer is moving upward.  The bottom panel shows
     the fractional differences caused by aberration for each of the
     five ring survey regions before (cyan points connected by dashed curve)
     and after (blue points) correcting for the sky mask window function.
     We also show the cosmic variance with $f_{\rm sky}=0.1$ with red dashed
     curves and the null case ($\Delta C_\ell/C_\ell=0$)with
     black solid curves.
     The magenta solid curves show our analytical approximation 
     of $\Delta C_\ell/C_\ell\approx -({\rm d}\ln C_\ell/{\rm d}\ln \ell)\beta\cos\theta$,
	  which reproduces well the simulation.
     A survey in the backward part of sky would yield residuals with the 
     opposite sign.
}
\label{fig:simCl_rings}
\end{figure*}

To illustrate the effect of aberration on the cut-sky power spectrum, 
we first consider a few toy masks with regular shapes in \reffig{simCl_rings}. 
In the top panel, the Mollweide-projected sky map is presented 
in a coordinate system where the direction of the observer's peculiar velocity 
aligns with the north pole. Each of the five ``ring''-shaped cut
skies 
considered here occupies a narrow range of
latitudes in the forward hemisphere (northern hemisphere) with
the same sky coverage $f_{\mathrm{sky}}=0.1$. 
However,  each is located at a different range of latitudes. 
We call them {\sf ring0} to {\sf ring4} from top (higher latitude) 
to bottom (lower latitude).
The five plots in the bottom panel show the fractional difference 
$\Delta C_\ell/C_\ell$ of the aberrated power spectra 
(with $\Delta \ell=50$ binning) from the unaberrated power spectra, 
for each of the five cut skies, respectively. Cyan dots show the fractional
differences between power spectra directly measured from the cut
skies while
blue dots (which lies on top of the cyan dots for small $\ell$) show the 
difference after deconvolving the cut-sky masking effect by inverting the
coupling matrix in the {\sc Master} algorithm \cite{Hivon:2001jp}.\footnote{%
Strictly speaking, this deconvolution only works when the 
two-point correlator is diagonal in spherical-harmonic space. 
Although the aberrated temperature map has a non-zero off-diagonal 
two-point correlation, the effect of these off-diagonal terms must
be small given that we were able to reproduce in
\reffigs{simCl_rings}{mCl_SPT} the expected 
residual after deconvolution.}
The black dashed curve following the cyan dots shows the moving average
of the residuals with $\Delta \ell=50$.
In the bottom plots, for reference, 
we also show the cosmic variance for $f_{\rm sky}=0.1$ as red dashed curves,
and indicate the un-biased case ($\Delta C_\ell/C_\ell=0$) with
black solid curves.

The residual plots in \reffig{simCl_rings} show that each cut-sky power 
spectrum is enhanced, $\Delta C_\ell/C_\ell>0$, by aberration,
with oscillatory features as a function of $\ell$.
We find that the oscillatory features in $\Delta C_\ell/C_\ell$
is in phase
with the slope of the power spectrum ${\rm d}\ln C_\ell/{\rm d}\ln\ell$.
Furthermore, the enhancement is largest (and exceeds the cosmic variance 
at each $\ell$) when the cut
sky is nearest to the north pole ({\sf ring0}) and hence has the
maximum asymmetry about the equator. The enhancement is
decreasingly prominent when the cut sky moves to lower
latitudes, and is least significant for {\sf ring4}. If the cut sky is
located in the backward hemisphere, aberration would instead
suppress the power spectrum.

This can be understood as follows:  Aberration rescales the size
of hot/cold spots as 
described in \refeq{Deltatheta_aberration}. 
In spherical-harmonic space, this results to the angle-dependent rescaling 
of the multipole moment. For a thin ring centered around latitude 
$\pi/2-\theta$, aberration rescales the multipole moment as 
\be
\ell \to \ell' = \ell \gamma\left(1 + \beta\cos\theta\right),
\label{eq:rescale_ell}
\ee
which induces a fractional change,
\be
\frac{\Delta C_\ell}{C_\ell}
= - \frac{\mathrm{d}\ln C_\ell}{{\rm d}\ln\ell}\beta\cos\theta
+ \mathcal{O}(\beta^2),
\label{eq:ring_prediction}
\ee
in the angular power spectrum.  We plot the theoretical
prediction as a magenta curve in 
\reffig{simCl_rings} which shows an excellent match between 
\refeq{ring_prediction} and the simulation (blue dots).

\begin{figure*}[htbp]
\centering
\includegraphics[width=0.49\textwidth]{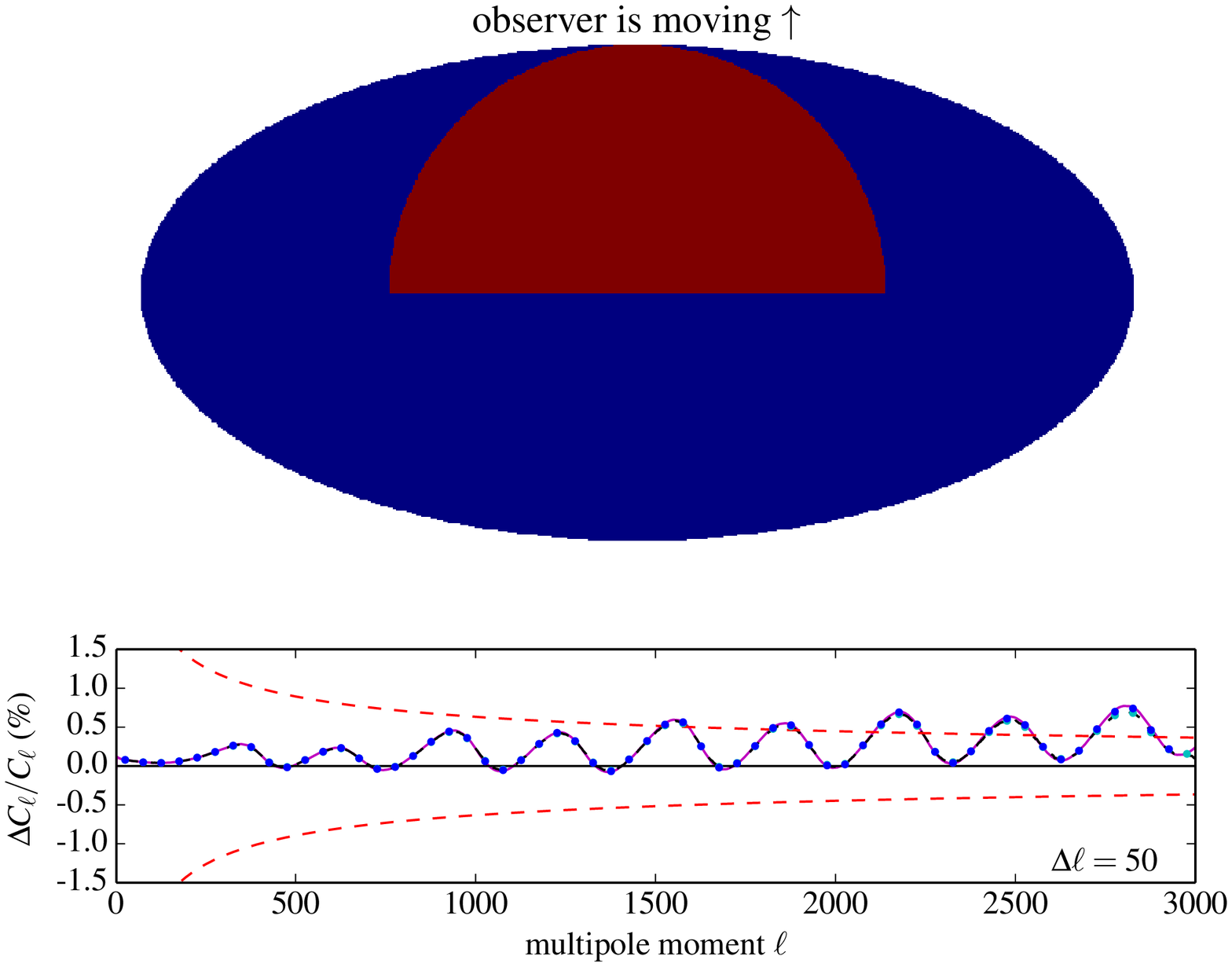}
\includegraphics[width=0.49\textwidth]{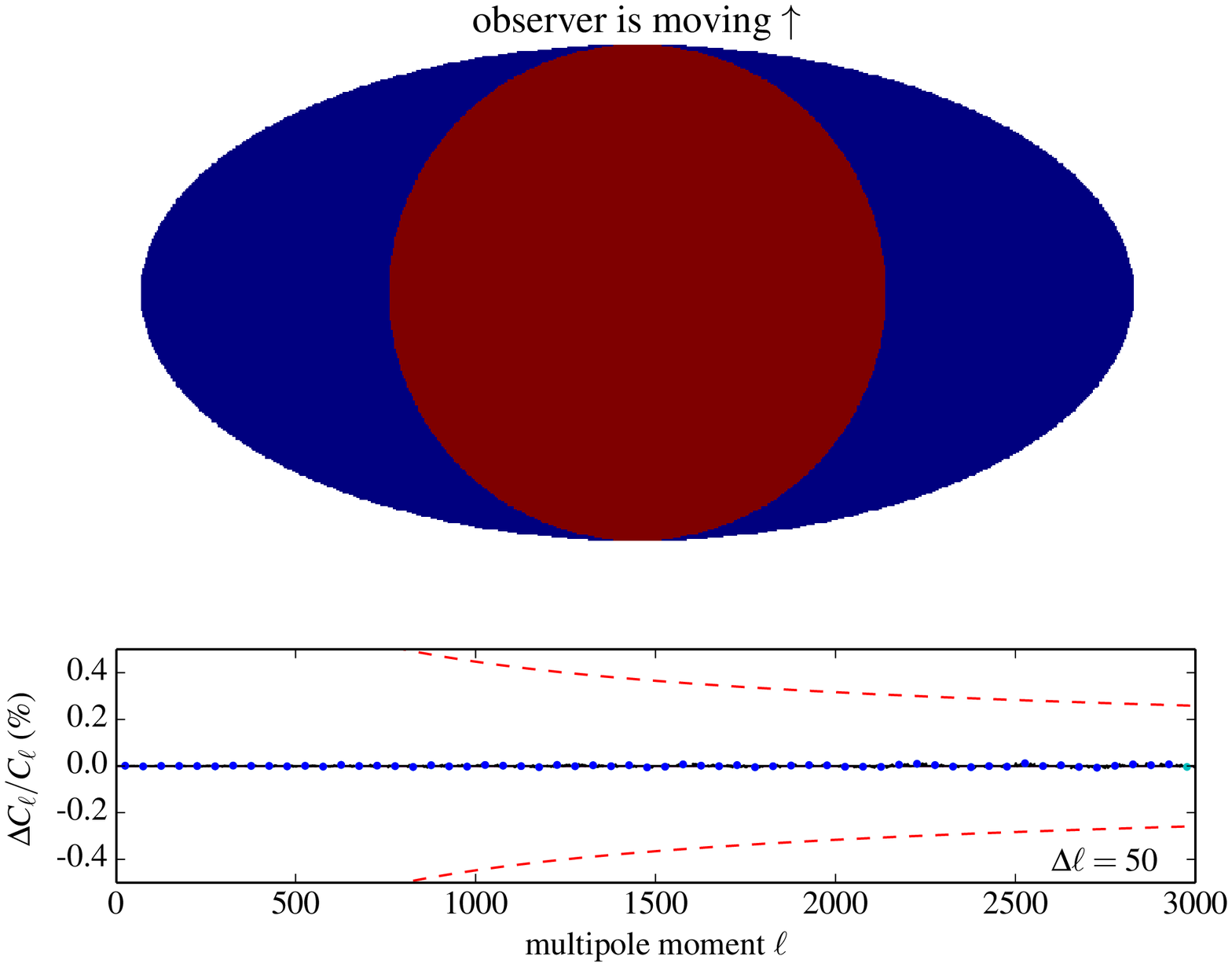}
\caption{Shape of two survey regions that cover only half the azimuthal
     regions. \textit{Left:} surveying only the forward hemisphere, and 
     \textit{right:} surveying the entire range of polar angles.
     For both regions, the bottom panels show the residual of the observed
	  power spectrum.  The curves and symbols are the same
          as in \reffig{simCl_rings}.  
     Note that $\left<\cos\theta\right>=0.5$ and
     $\left<\cos\theta\right>=0$, respectively,
     for the left and right panels.  Thus, the residuals in the 
     symmetric survey region (right panel) suppressed by a
     factor of $\beta$ relative to the asymmetric survey in the
     left panel.}
\label{fig:simCl_vhalf}
\end{figure*}

In \reffig{simCl_vhalf}, we consider two other scenarios: one
has half of the forward hemisphere 
(covering azimuthal angles $0\leq\phi\leq 180^\circ$) 
being surveyed ($f_{\mathrm{sky}}=0.25$), and the other spans the full range 
of the polar angle with the same azimuthal angle coverage.
Again, the maximum of $\simeq 0.8\%$ enhancement in the aberrated power
spectrum is seen for the forward case and at large $\ell$s the bias
dominates over the cosmic variance. For the right-hand-side region, where 
the survey area encloses both forward and backward regions in a symmetric 
manner, the aberrated power spectrum shows only very small
differences from the unaberrated power spectrum.

What about the theory prediction for these extended regions?
For the extended survey region which is defined by the masking function 
$M(\theta,\phi)$ (which takes $1$ for the observed region, and $0$ otherwise), 
we superpose the residuals \refeq{ring_prediction} from many thin stripes
as 
\ba
\frac{\Delta C_\ell}{C_\ell}
\approx&
- \frac{\beta}{4\pi f_{\rm sky}} 
\frac{{\rm d}\ln C_\ell}{{\rm d}\ln \ell}
\!\int_{0}^\pi\! {\rm d}(\cos\theta) \cos\theta \int_0^{2\pi} 
{\rm d}\phi M(\theta,\phi)
\vs
=&
-\frac{{\rm d}\ln C_\ell}{{\rm d}\ln \ell}\beta \left<\cos\theta\right>
\label{eq:DClovCl_linear}
\ea
where $\left<\cos\theta\right>$ is the area-averaged mean of the polar cosine.
Note that the residual does not depend on the azimuthal angle because 
aberration is symmetric in the azimuthal angle.
The magenta curve in the left bottom panel of \reffig{simCl_vhalf} shows the 
prediction from \refeq{DClovCl_linear} ($\left<\cos\theta\right>=0.5$),
which provides excellent agreement with the simulation.
We also confirm that the linear approximation in \refeq{DClovCl_linear} models
the residual accurately for the realistic cut skies of Planck, SPT and ACT
in \reffigs{mCl_Planck}{mCl_SPT}.

One of the most important implication of \refeq{DClovCl_linear} is that 
the linear order ($\propto\beta$) aberration effect vanishes
for surveys with forward/backward symmetry. This is what we see in the
right panel of \reffig{simCl_vhalf}. 
The tiny, residual $\Delta C_\ell/C_\ell$, with an
oscillatory behavior about zero, is negligible compared with
cosmic variance. This is expected, because, in linear order, 
the enhancement of power from the forward direction cancels out the 
suppression of power from the backward direction.  The net
effect is thus suppressed by a factor $\beta$ compared to the 
asymmetric cases:
\be
\frac{\Delta C_\ell}{C_\ell}
\approx
\frac{\beta^2}{2}
\left[
\frac{{\rm d}\ln C_\ell}{{\rm d}\ln \ell}
+
\frac{1}{C_\ell}\frac{{\rm d}^2C_\ell}{{\rm d}\ell^2}\ell^2
\left<\cos^2\theta\right>
\right]
\label{eq:DClovCl_second}
\ee
Here, $\left<\cos^2\theta\right>$ is the area average of the square of the 
polar cosine. We compare \refeq{DClovCl_second} with the full sky power 
residual ($\left<\cos^2\theta\right>=1/3$)
in the bottom panel of \reffig{simCl_fullsky}, and find that it
captures  the overall shape of the full-sky residual very accurately.

\subsection{Planck}
\begin{figure*}[htbp]
\includegraphics[width=0.49\textwidth]{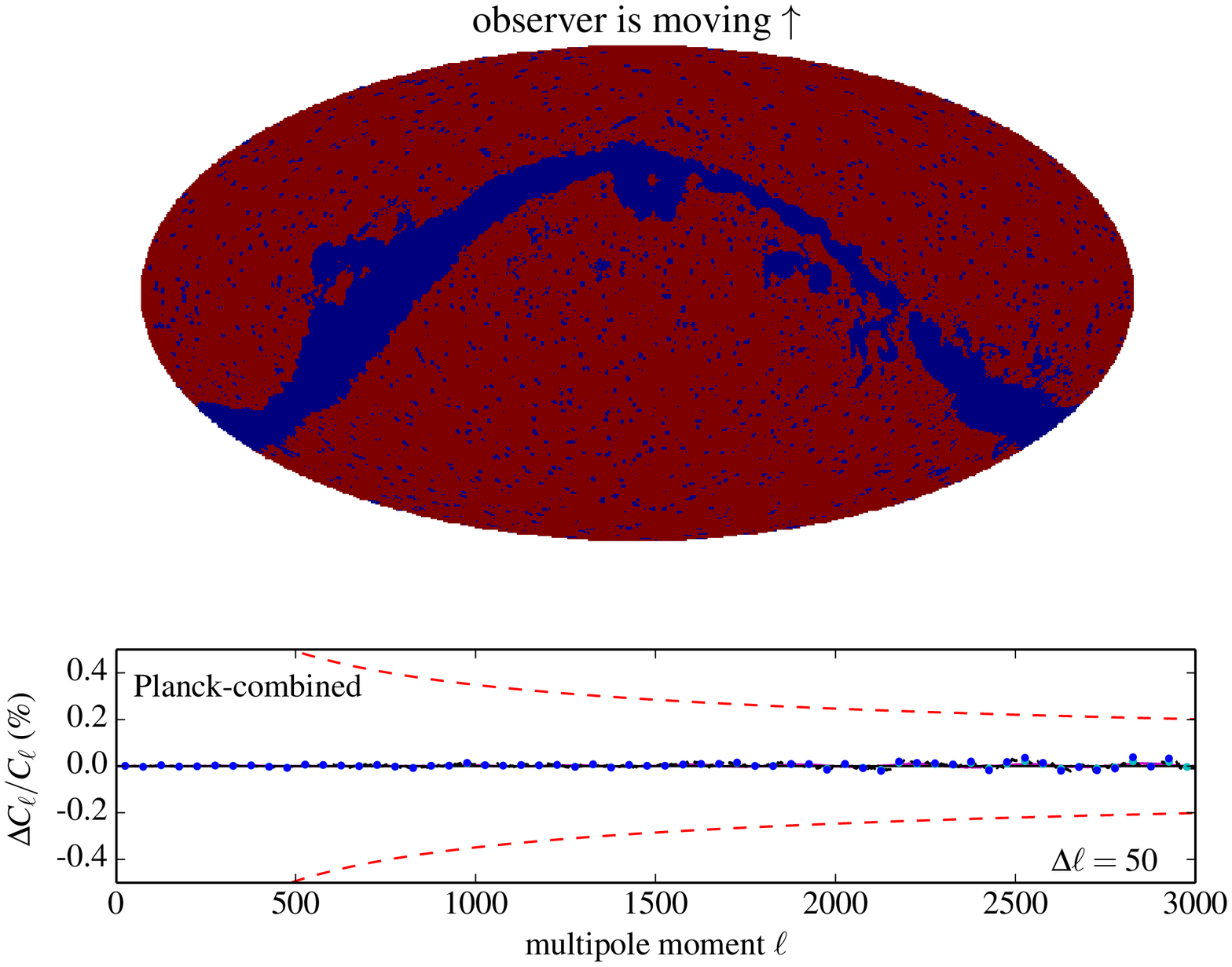}
\includegraphics[width=0.49\textwidth]{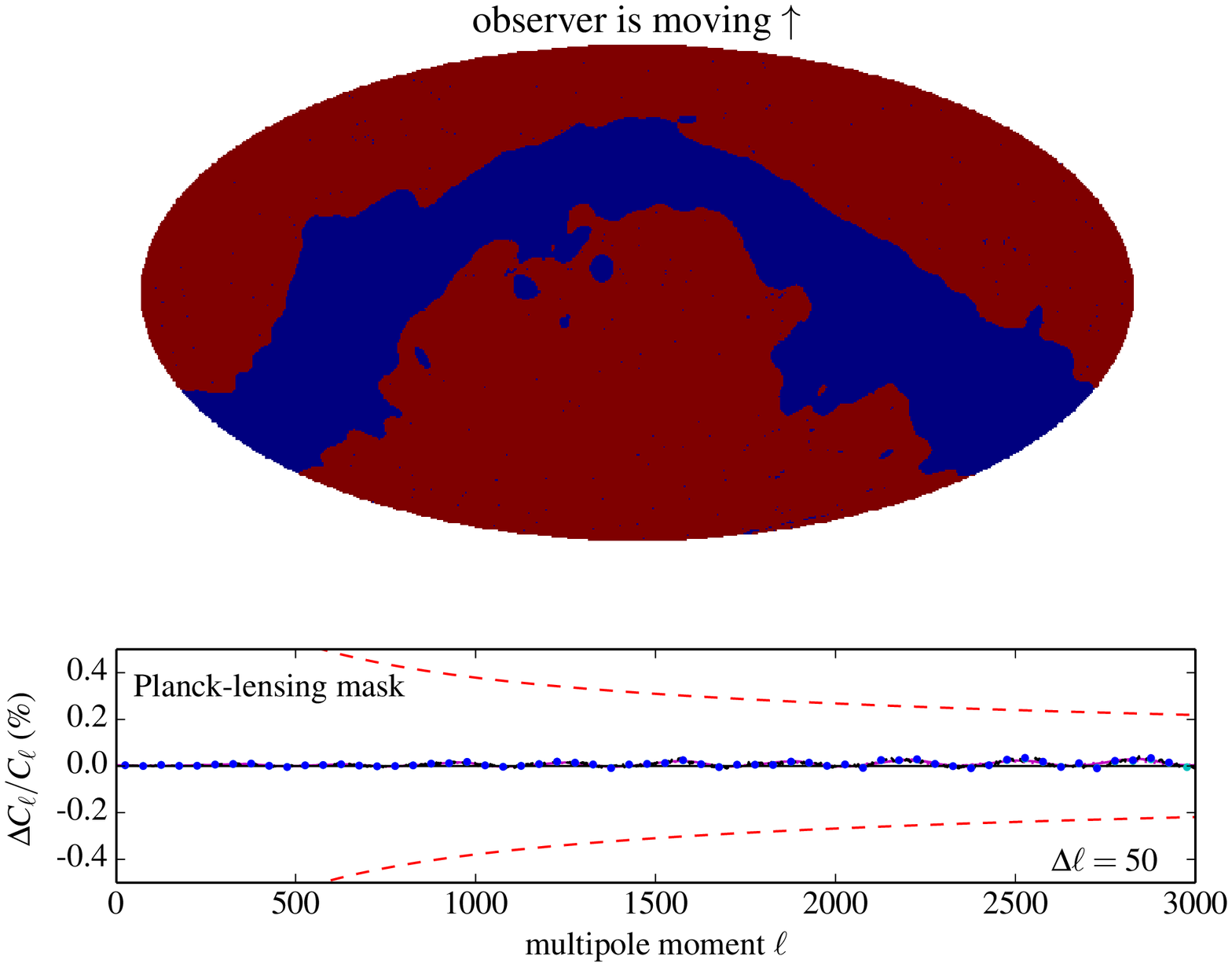}
\includegraphics[width=0.49\textwidth]{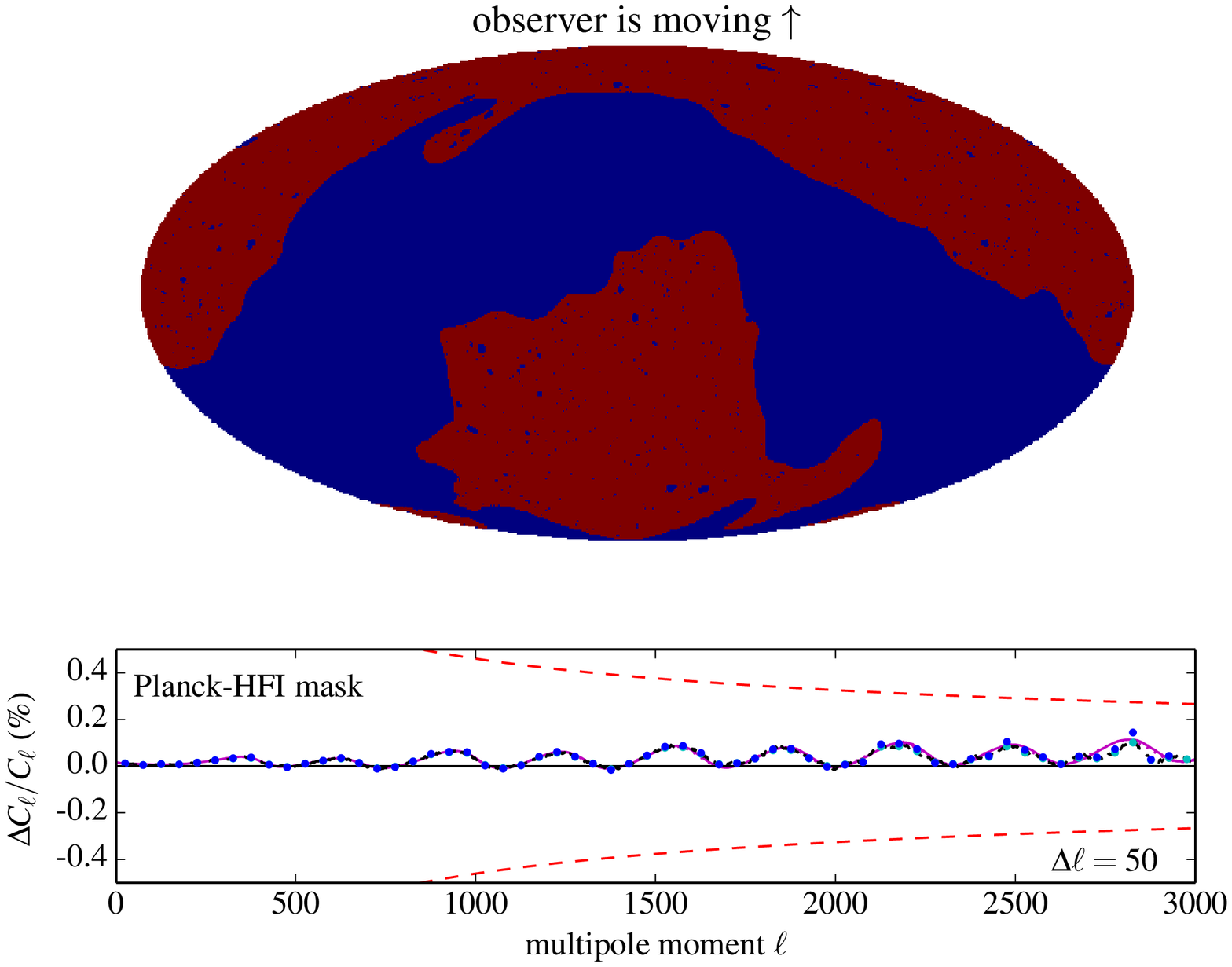}
\includegraphics[width=0.49\textwidth]{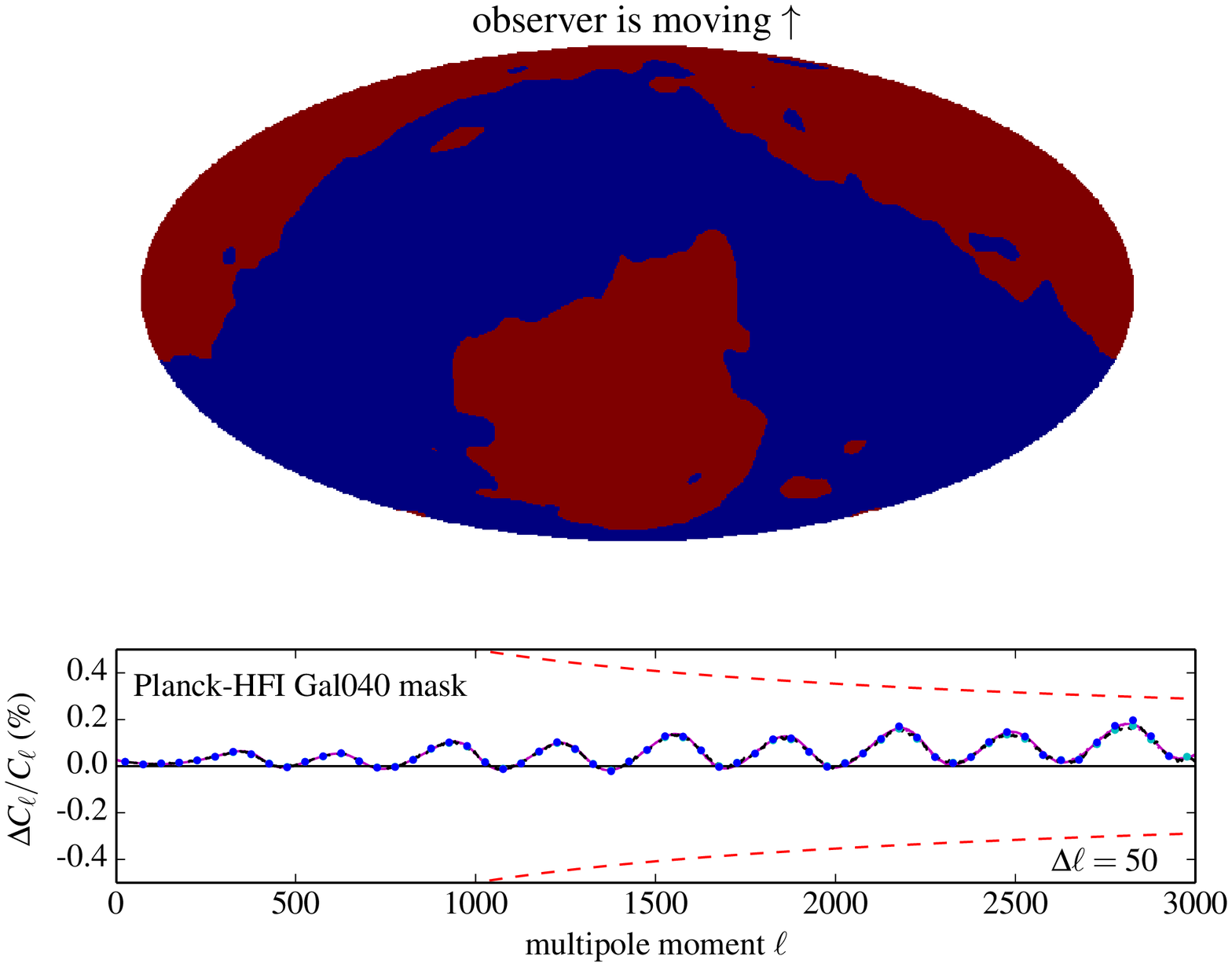}
\caption{The shape of the mask (\textit{top panel}) and the
     residual of power spectrum (\textit{bottom panel}) for four different
     masks from Planck satellite: Combined mask (\textit{top left}), 
     Lensing mask (\textit{top right}), HFI mask (\textit{bottom left}) and 
     HFI Gal040 mask (\textit{bottom right}). 
     The mask is drawn in coordinates where the observer is moving upward.  
     The bottom panel of each figure shows the residual effect of aberration 
     before (cyan points connected by dashed curve) and after (blue points) the
     sky-mask deconvolution along with the cosmic-variance error
     (red dashed curve) with 
     $f_{\rm sky}=0.823$, $0.697$, $0.471$ and $0.400$ for 
     Combined, lensing, HFI, HFI Gal040 mask, respectively.
     Each point shows the binned average with the indicated
     width, and the magenta curve shows the prediction from 
     the linear theory in \refeq{DClovCl_linear}
     with $\left<\cos\theta\right>=0.002$ (combined mask), 
     $0.015$ (lensing mask), $0.07$ (HFI mask) and  $0.114$ (HFI Gal040 mask).
     Note that aberration residual is different for different masks.
     If cosmic variance dominates the error budget down to $\ell=3000$,
     aberration biases the amplitude of the angular power spectrum
     to $0.3\sigma$, $0.4\sigma$, $1\sigma$, and $1.6\sigma$ for 
	  combined, lensing, HFI and HFI Gal040 mask, respectively.
}
\label{fig:mCl_Planck}
\end{figure*}

As shown in \reffig{mCl_Planck}, using the same coordinates as before, 
Planck masks out the Galactic disk to minimize microwave contamination
from our Galaxy, thus leaving a fraction
$f_{\mathrm{sky}}\simeq0.823$, $0.697$, $0.471$  and $040$ of the sky for 
combined mask, lensing mask, HFI mask and HFI Gal040 mask, respectively,
for the CMB analysis.
The geometry of the sky mask,
which can be approximated by a narrow band around zero Galactic
latitude, is not symmetric about the boost equator.
The degree of forward/backward hemispherical asymmetry depends 
significantly on the choice of the mask. While the combined mask has roughly 
same area in the forward hemisphere as in the backward hemisphere
($\left<\cos\theta\right>=0.002$), 
the other masks show larger asymmetries with 
$\left<\cos\theta\right>=0.015$, $0.07$, and $0.114$ for the lensing mask, 
HFI mask and HFI Gal040 mask, respectively.

Consequently, it can be seen from the accompanying plot in 
\reffig{mCl_Planck} that aberration for Planck with a Galactic
mask changes dramatically from one mask to another.
The combined mask has negligible impact across the whole range of the power spectrum, but the HFI masks show residuals comparable to the cosmic-variance error on small angular scales ($\ell\gtrsim2000$).  Thus, the bias in the deduced power spectrum is at the $1\sigma$ level, if the error budget is dominated by cosmic variance.  Therefore, the effect of aberration must be corrected differently for the different masks to achieve an unbiased measurement of the temperature power spectrum in the CMB rest frame.  As the Galactic masks for different frequency channels show different degrees of backward/forward asymmetry (thus different aberration effects), the aberration, if neglected, may create some tension among power spectra estimated from different frequency channels.

\subsection{SPT, ACT-S, ACT-E}
\begin{figure*}[htbp]
\includegraphics[width=0.49\textwidth]{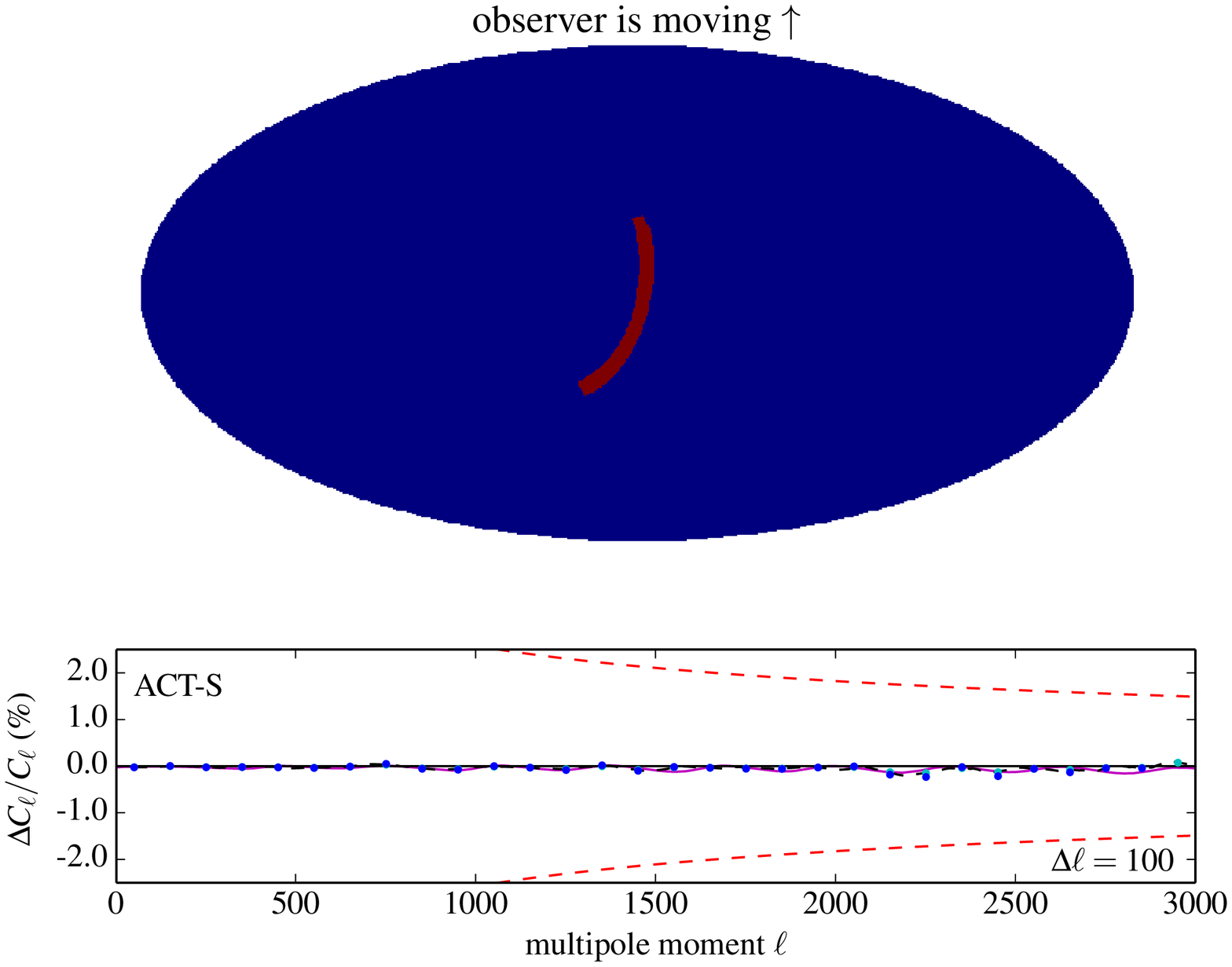}
\includegraphics[width=0.49\textwidth]{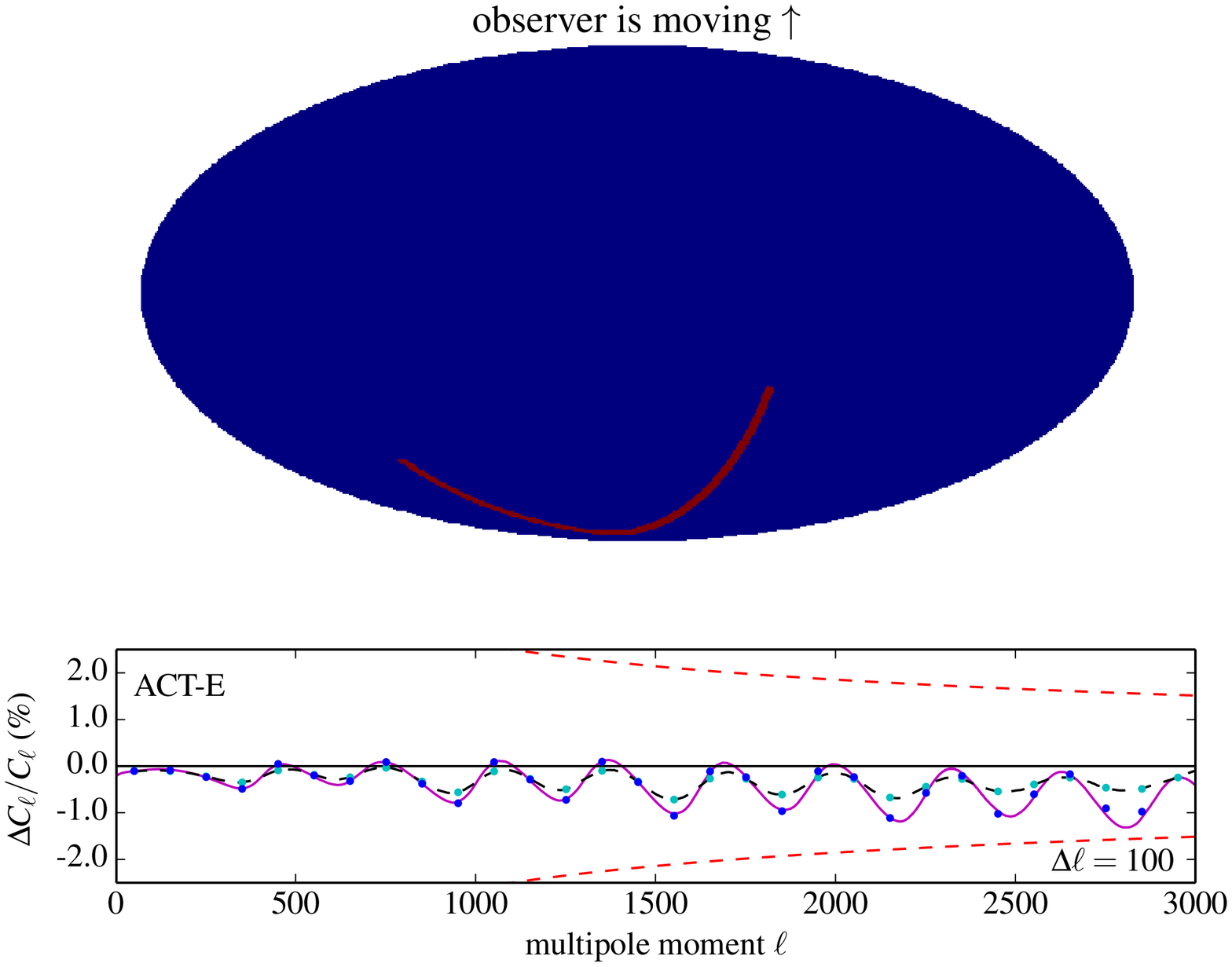}
\caption{The same as \reffig{mCl_Planck}, but for two of the ACT regions:
	  ACT-S(\textit{left}) and ACT-E(\textit{right}).
     The cosmic-variance error (red dashed curve) is calculated with 
	  $\Delta \ell=100$ and $f_{\rm sky}=0.0075$, $0.0073$ for 
     ACT-S and ACT-E, respectively.
     The magenta curve shows the prediction from the linear theory 
     in \refeq{DClovCl_linear}
     with $\left<\cos\theta\right>=-0.18$ (ACT-S) and $-0.85$ (ACT-E).
     If cosmic variance dominates the error budget down to $\ell=3000$,
     aberration biases the amplitude of the angular power spectrum
     to $0.2\sigma$ and $1\sigma$ for ACT-S and ACT-E, respectively.
}
\label{fig:mCl_ACTs}
\end{figure*}

\begin{figure}[htbp]
\includegraphics[width=0.49\textwidth]{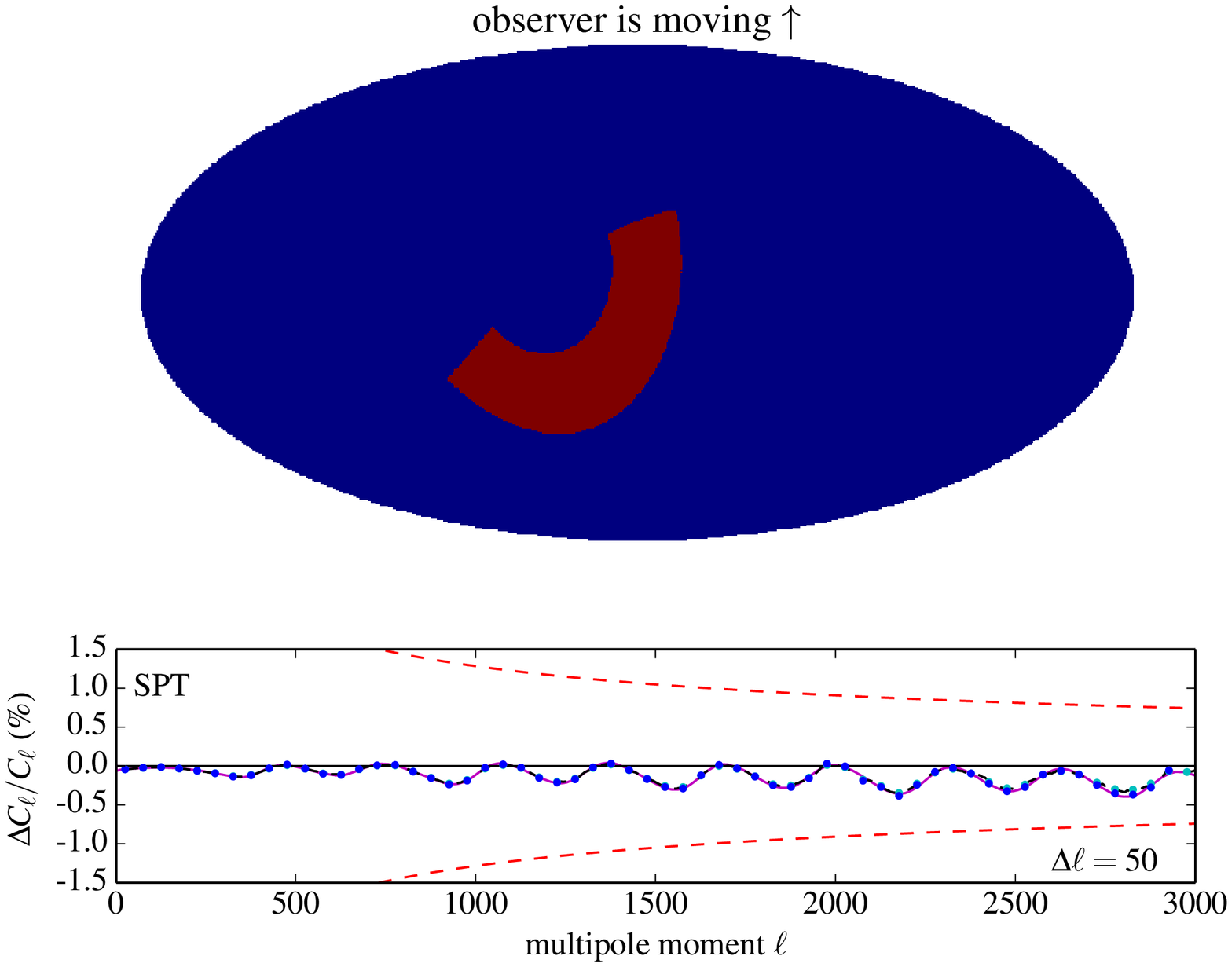}
\caption{The same as \reffig{mCl_Planck} and \reffig{mCl_ACTs}, but for 
	  SPT footprint.
     The cosmic-variance error (red dashed curve) is calculated with 
	  $\Delta \ell=50$ and $f_{\rm sky}=0.06$, and the magenta curve 
     shows the prediction from the linear theory in \refeq{DClovCl_linear}
     with $\left<\cos\theta\right>=-0.26$.
     If cosmic variance dominates the error budget down to $\ell=3000$,
     aberration biases the amplitude of the angular power spectrum
     sky-mask deconvolution along with the cosmic-variance error
     to $1.3\sigma$.
}
\label{fig:mCl_SPT}
\end{figure}

Aberration is also important for
ground-based experiments like ACT and SPT, where
the sky coverage is smaller and where the mask can be highly
asymmetric with respect to the aberration equator.

ACT has so far observed two long narrow strips across the
sky. In the left panel of \reffig{mCl_ACTs}, 
the southern strip, corresponding to the ACT-S mask, is shown
with a sky coverage
$f_{\mathrm{sky}}=0.0075$~\cite{Das:2013zf}. Coincidentally, the
southern strip is roughly symmetric about the aberration
equator, with the part in the backward hemisphere slightly
larger than that in the forward hemisphere. Therefore, the
cut-sky power spectrum for the ACT-S mask is only marginally suppressed
by aberration, with $|\Delta C_\ell/C_\ell|<0.25\%$. The
equatorial strip, or ACT-E mask, has a sky coverage of
$f_{\mathrm{sky}}=0.0073$, and lies completely in the backward
hemisphere~\cite{Das:2013zf}, as shown in the right panel of 
\reffig{mCl_ACTs}. As
a result, our simulations exhibit a more significant power
suppression from aberration, as large as $\simeq 1\%$ in
magnitude.

Also, as shown in \reffig{mCl_SPT}, SPT has surveyed a
region in the sky with $f_{\mathrm{sky}}=0.06$. It can be seen
that a much larger fraction of the survey region
lies in the backward hemisphere, which leads to as much
as an $0.4\%$ suppression of the power spectrum for
$\ell\gtrsim 1000$ caused by the observer's peculiar
velocity. 

Therefore, our simulations unambiguously
demonstrate that both the ACT and SPT experiments suffer from
systematic power suppression on small scales from aberration.
As the Figures indicate, the suppression is small, in each
$\ell$ band plotted, compared with the statistical error.
Still, the systematic bias induced by aberration in the {\it
complete} data set, and thus on the inferred cosmological
parameters, is in fact significant.  For example, the SPT power
spectrum measured to $\ell_{\rm max}\simeq3000$ from 6\% of the
sky is obtained from $\simeq 5.6\times10^5$ modes, implying a
cosmic-variance error of $\simeq 0.0013$ on the overall
amplitude.  Thus, the systematic suppression in SPT due to
aberration may be as large as a few-$\sigma$ effect and cannot be ignored.
Given the smaller suppression and smaller sky coverage in ACT-S,
the effects of aberration are smaller compared with the
statistical error.  However, for ACT-E, the systematic
suppression due to aberration may be a roughly $\simeq1\sigma$
effect.  We surmise that aberration may be at least partially
responsible for the small tension between cosmological
parameters inferred from ACT, SPT, and Planck\cite{DiValentino:2013mt}.

\subsection{Aberration effect on cosmological parameters}
\begin{figure}[htbp]
\includegraphics[width=0.49\textwidth]{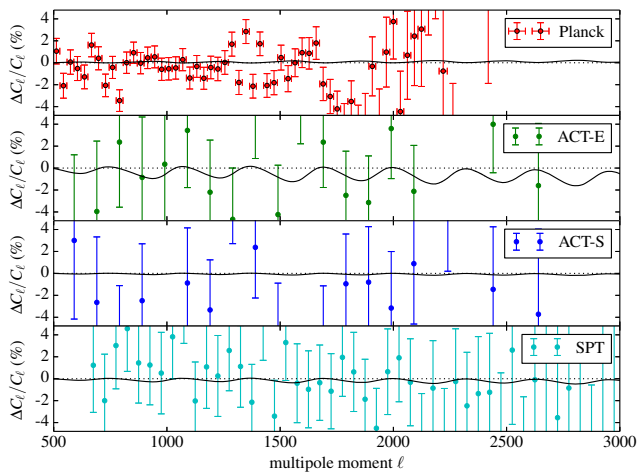}
\caption{The power-spectrum residuals for Planck, ACT-E, ACT-S,
      and SPT 
     (from \textit{top} to \textit{bottom}) with respect to the best-fitting
     power spectrum from combining Planck, lensing, WMAP
     polarization and high $\ell$ 
     (ACT and SPT). The data points are the measured binned power spectrum, and 
     the error bars include both cosmic variance and instrumental noise.
     The expected aberration residual with $\beta=0.00123$
     is shown as the black solid curve.
     With the error bars used in this plot, aberration biases
     the amplitude of the angular power spectrum
     by $0.47\sigma$, $0.07\sigma$, $0.64\sigma$, and $0.34\sigma$ for 
	  Planck (Gal040 mask), ACTS, ACTE and SPT, respectively.
}
\label{fig:mCl_data}
\end{figure}

In the previous Sections, we have shown that the effect of aberration on
the measurement of the temperature power spectrum on part of the sky can be 
accurately modeled by the simple re-scaling of multipole moments in 
\refeq{rescale_ell}.
Therefore, ignoring this re-scaling due to aberration biases all of the 
`horizontal' information encoded in the temperature power spectrum. 
For example, ignoring aberration would shift 
the sound-horizon angle $\theta_*$ by 
$\Delta\theta_*/\theta_* = -\Delta\ell/\ell\simeq-\beta\left<\cos\theta\right>$
which is $-0.014\%$ with the Planck Gal040 mask, and goes up to 
$0.03\%$ and $0.1\%$ with, respectively, the SPT and ACT-E survey footprints.
The level of bias in measuring $\theta_*$ is comparable to the 
reported 1-$\sigma$ ($68\%$ confidence level) 
range of measuring $\theta_*$ ($\sim0.06\%$) from the Planck collaboration
\cite{Ade:2013zuv}\footnote{Planck collaboration has also noted this shift of 
$\theta_*$ in footnote 13 (page 9) of \cite{Ade:2013zuv} and page 2 of 
\cite{Aghanim:2013suk}.
}.

How does this shift affect the cosmological parameters?
A full answer to this question requires an in-depth analysis of the likelihood 
function in the parameter space. We simplify the situation by
fixing parameters which determine the shape of the power psectrum,
$\{\Omega_mh^2$, $\Omega_bh^2$, $\tau$, $n_s$, $A_{s}\}$ so that the shift
in the sound-horizon angle $\theta_*$ mainly causes bias in 
the angular-diameter distance $D_A(z_*)$ to the last-scattering surface. 
Assuming a flat universe, this leaves us with two parameters: 
the Hubble parameter $H_0= 100h~\mathrm{Mpc}/\mathrm{km}/\mathrm{s}$ 
and dark-energy equation of state $w_{\rm de}$\footnote{
Fixing $\Omega_mh^2$ in flat universe also fixes dark energy density parameter
$\Omega_{\mathrm de}$ for a given $h$.}.
With the best-fitting $\Lambda$CDM parameters 
(``Planck+lensing+WP+highL'' column in the Table 5 of \cite{Ade:2013zuv}),
we find that aberration induces an $0.072\%$ bias on $H_0$ 
(that moves maximum likelihood value from $h=0.6794$ to $h=0.6799$)
and $-0.16\%$ bias on $w$ for Planck HFI (Gal040) mask,
and it goes up to $-0.15\%$ ($H_0$), $0.34\%$ ($w$) for SPT, 
and $-0.51\%$ ($H_0$), $1.14\%$ ($w$) for ACT-E.
Therefore, it is important to correct for aberration in order to
acheive a percent level accuracy on parameters such as $H_0$ and $w$.
Moreover, when combining parameters from different surveys, 
ignoring aberration would enhance the tension as cosmological parameters 
from Planck and ACT/SPT are biased toward opposite directions.

Finally, to quantify more precisely the magnitude of the effect in
current data set, \reffig{mCl_data} shows power-spectrum
residuals with respect to the best-fitting $\Lambda$CDM model
(``Planck+lensing+WP+highL'' column in the Table 5 of
\cite{Ade:2013zuv}) for the four CMB surveys we consider here:
Planck \cite{Planck:2013kta}, ACT-E, ACT-S, and SPT 
\cite{Calabrese:2013jyk}. The error bars here include not
only cosmic variance, but also the instrumental noise.  With the
effects of instrumental noise for the existing data included,
the magnitude of the effect of aberration relative to the
statistical error is reduced. The systematic bias induced by
aberration in the amplitude of the power spectrum becomes
$0.47\sigma$
$0.64\sigma$ and $0.34\sigma$ for Planck HFI (Gal040 mask), ACT-E and SPT, 
respectively. This is smaller than what would be inferred considering only
cosmic variance, but still not negligible.

We furthermore note that the effects of aberration may have a more profound
impact on power-spectrum and cosmological-parameter measurements
from future experiments that survey larger regions of the sky
(especially those that are skewed toward one of the boost
poles).  It will also become more significant for forthcoming
experiments that include polarization, an issue we address in a
forthcoming publication \cite{Dai:future}.
Also, aberration might affect  higher order statistics of the CMB both on the full-sky and for the cut-sky in subtle ways that should be considered more carefully.

\section{Conclusion}\label{sec:conclusion}

While aberration affects full-sky measurements of the power
spectrum only at order $\beta^2$ [\refeq{DClovCl_second}], 
the effect on the power
spectrum inferred from maps with partial-sky coverage arise at
linear order in $\beta$ [\refeq{DClovCl_linear}]
and may thus be much larger.  The
effects of aberration are magnified further as a consequence of
the steep falloff of the power spectrum  ($C_\ell \propto
\ell^{-7}$) at the high $\ell$ probed by several current and
forthcoming CMB experiments.  In total, the effect can
constitute a systematic bias as large as $\simeq1\%$, considerably
larger than statistical errors in current measurements, in
measurements of the power spectrum.  It must therefore be
explored in detail.

We have developed a novel formalism to account for the effects of
aberration on measurements of the CMB temperature power spectrum
from maps with partial sky coverage.  Our analysis improves upon
prior work by going to higher orders in $\beta$, thus extending
the validity of analytic calculations to multipole moments $\ell
\gtrsim \beta^{-1}\simeq800$.  Our harmonic-space approach to
de-boosting also provides a more effective and computationally
efficient way to deal with the effects of window functions and
pixelization than the real-space approach explored in prior work.

We then used these new algorithms to explore in detail the
effects of aberration on Planck, ACT, and SPT.  
We conclude that the effect of aberration for Planck depends strongly
on the choice of the mask, and the mask used for power spectrum analysis 
of data with HFI (Gal040) shows $\simeq\sigma$ level changes due to 
aberration. 
More importantly, as the effect of aberration varies from one mask to another,
unbiased estimation of temperature power spectrum would require 
cleaning the aberration effect before combining the power spectra 
from different frequency channels.
We also conclude that the systematic bias in current SPT
data affects measurements of the power spectrum at the
$\simeq\sigma$ level and thus cannot be ignored.  The effects in
current ACT-S data are negligible, but those in ACT-E arise at
the $\simeq\sigma$ level.  We surmise that aberration may be
responsible for part of the small tension between power spectra inferred
from SPT, ACT, and Planck, and thus also for the values of
cosmological parameters inferred from these experiments.

We also note that the magnitude of the effects of aberration,
relative to the statistical error, may become larger for future
measurements that survey larger regions of the sky, especially
for those that are aligned or anti-aligned with our peculiar
velocity with respect to the CMB rest frame.  Aberration will
also become more important with forthcoming experiments in which
the statistical power is extended with measurements of the
polarization.  A forthcoming paper \cite{Dai:future} will deal
with aberration effects on the polarization.

Finally, while the focus in our discussion has been on
measurements of the {\it primordial} power spectrum, and the
values of cosmological parameters inferred, aberration affects
the measurement of {\it all} fluctuations.  This includes CMB
fluctuations from secondary extragalactic sources, and it will
also include measurements of other cosmic backgrounds 
such as fluctuations of the 21-cm background. 
High-resolution measurement of weak lensing may also be affected
by aberration.
As the analytical calculations in \refeqs{DClovCl_linear}{DClovCl_second}
come solely from relativistic beaming and does not depend on the specific 
shape of the energy spectrum, it should also be useful for analyzing the 
power spectra for these other maps.

Our harmonic-space approach to deboosting will also be useful for 
measurements of higher-order statistics as well.  
Without them, estimators for the bispectrum and trispectrum may be 
affected by aberration.

\begin{acknowledgments}
The authors acknowledge 
Anthony Challinor, 
Duncan Hanson,
Gilbert Holder,
Arthur Kosowsky,
Eiichiro Komatsu,
David Larson, 
Antony Lewis
and Aditya Rotti
for useful discussions, and Antony Lewis for pointing out the 
Planck mask used for the power spectrum analysis of HFI data.
Some of the results in this paper have been derived using the 
{\sc Healpix} \cite{Gorski:2004by} package.
This work was supported by DoE SC-0008108, NASA NNX12AE86G, and
NSF 0244990. MK acknowledges the hospitality of the Aspen Center
for Physics, supported in part by the National Science Foundation
under Grant No. PHYS-1066293.
LD is supported by the William Gardner Fellowship.
\end{acknowledgments}

\begin{figure}[htbp]
\includegraphics[width=0.52\textwidth]{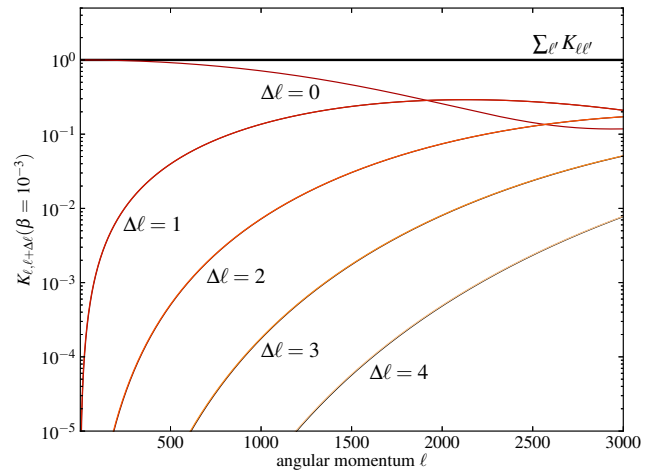}
\caption{The power transfer matrix, $K_{\ell\ell'}(\beta)$
     [\refeq{def_Kllp}], for $\beta=10^{-3}$. The results for
     $\ell>\ell'$ and $\ell<\ell'$ are the same to within the thickness of
     the curve.  For smaller angular scales ($\ell>200$), more
     than $1\%$ of the temperature fluctuation is transferred
     to adjacent multipoles, and power leakage reaches about
     $30\%$ for $\ell\simeq1000$. At $\ell=3000$, only about
     $12\%$ of the power is left at the same multipole
     ($\Delta l=0$).  The {\it total} curve at the top shows
     $\sum_{\ell'}K_{\ell\ell'}$.}
\label{fig:Kllp}
\end{figure}

\appendix
\section{Unitarity of the aberration kernel and approximate
power conservation}\label{app:unitarity}

Ref.~\cite{Challinor:2002zh} derived a relation, $C_\ell^{\rm
obs} = C_\ell ( 1 + 4 \beta^2 +\cdots)$, implying that the 
full-sky power spectrum for the boosted map differs by no more
than $\simeq 4\beta^2 \simeq6\times 10^{-6}$ from that for the
original map.  This analytic result does not necessarily hold,
however, at $\ell\gtrsim800$, as discussed above.  Numerical
evaluation of power spectra on boosted maps have since then
confirmed \cite{Yoho:2012am}, though, that the full-sky power
spectrum is very nearly unchanged.  Here we show that this
result follows directly from unitarity of the aberration kernel.

From \refeq{two-pt-correlator}, we calculate the observed
angular power spectrum,
\ba
C_\ell^{({\rm obs})}
=&
\frac{1}{2\ell+1}
\sum_{m}
\left<
\left|a_{\ell m}^{({\rm obs})}\right|^2
\right>
\vs
=&
\frac{1}{2\ell+1}
\sum_{\ell'm}
\left[\mathcal{K}_{\ell m}^{\ell'm}(\beta)\right]^2
C_{\ell'} 
= \sum_{\ell'} K_{\ell\ell'}(\beta) C_{\ell'},
\ea
where we defined the power transfer matrix 
\be
K_{\ell\ell'}(\beta) \equiv \frac{1}{2\ell+1}\sum_{m}
\left|\mathcal{K}_{\ell m}^{\ell'm}(\beta)\right|^2
\label{eq:def_Kllp}.
\ee

Unitarity of the aberration kernel arises because an aberration
followed by an inverse aberration (an aberration with negative
velocity $-\boldsymbol\beta$) should lead to the original map.
From \refeq{kappa}, we find \cite{Chluba:2011zh},
\citep[e.g.,][]{Challinor:2002zh,Chluba:2011zh}
\be
\mathcal{K}_{\ell m}^{\ell'm}(\beta)
=
\mathcal{K}_{\ell'm}^{\ell m}(-\beta).
\ee
Then, the aberration followed by inverse-aberration is
\ba
a_{\ell'' m}''
=&
\sum_{\ell'} 
\mathcal{K}_{\ell''m}^{\ell'm}(-\beta)
\left[
\sum_\ell
\mathcal{K}_{\ell'm}^{\ell m}(\beta) a_{\ell m}
\right]
\vs
=&
\sum_\ell
\sum_{\ell'} 
\mathcal{K}_{\ell'm}^{\ell''m}(\beta)
\mathcal{K}_{\ell'm}^{\ell m}(\beta) 
a_{\ell m} = a_{\ell m},
\ea 
which completes the proof of the unitarity,
\be
\sum_{\ell'} 
\mathcal{K}_{\ell'm}^{\ell''m}(\beta)
\mathcal{K}_{\ell'm}^{\ell m}(\beta) 
=
\delta_{\ell'' \ell}.
\label{eq:unitarity}
\ee

This result implies that the power transfer matrix satisfies
$\sum_{\ell'} K_{\ell\ell'}=1$.  Since, roughly speaking,
aberration shifts multipole moments $\ell \to \ell(1\pm\beta)$
in the forward/backward directions, the effect of aberration on
a full-sky map is to smear the power over a range $\Delta \ell
\simeq \beta\ell$.  Numerical evaluation of the kernel, shown in
\reffig{Kllp}, verifies that the support of the power transfer
matrix $K_{ \ell,\ell+\Delta\ell}$ is limited to $\Delta\ell
\lesssim \beta\ell$.

Smearing of the power spectrum $C_\ell$ over a range $\beta\ell$
then leads to a power change $\simeq (\partial^2 C_{\ell}/\partial
\ell^2) \beta^2\ell^2$.  Thus, the smallness of the shift in the
full-sky power spectrum at $1000\lesssim \ell \lesssim3000$ is
seen to be a consequence of the fact that $\beta\ell \simeq 1$ is
small compared with the spacing $\Delta\ell\simeq 200$ between the
acoustic peaks.

\end{document}